\documentstyle[multicol,aps,prl]{revtex}

\begin{document}

\begin{multicols}{2}

\noindent
{\bf Usov Replies:}
In the recent Letter [1] I have shown that the Coulomb barrier at 
the quark surface of a hot strange star may be a powerful source 
of $e^+e^-$ pairs. These are created in a very strong electric 
field, $E\gg E_{\rm cr}\simeq 1.3\times 10^{16}$ V cm$^{-1}$, 
of the barrier and flow away from the stellar surface. 
The flux of $e^+e^-$ pairs from the quark surface {\it with a given 
temperature} was calculated in Ref. [1]. In the Comment [2] Mitra claims 
that the process of pair creation rapidly leads to quenching 
of the surface electric field, and in a steady state, the value of $E$ 
at the quark surface is well below the critical value, $E\ll E_{\rm cr}$.
However, this claim is incorrect because the electric field of the
Coulomb barrier at the quark surface cannot be quenched for 
a long time in principle. The point is that there are some electrons 
in quark matter to neutralize the electric charge [3,4].
The electrons are bound to the quark matter by the electromagnetic 
interaction and not by the strong force, and the  Coulomb 
barrier with a very strong electric field has to exist at the quark 
surface to prevent the electrons from escaping to 
infinity [3,4]. Even if the Coulomb barrier is quenched at some 
moment, the electron distribution would extend very fast up to the 
distance $\Delta R\sim 10^{-10}$ cm above the quark surface, 
and the Coulomb barrier would be restored
in $\sim\Delta R/c\sim 3\times 10^{-21}$ s. Here, it is taken
into account that the Fermi energy of electrons in quark matter is
much higher than their rest energy, $\varepsilon_{\rm F}\gg mc^2$, and 
the mean velocity of electrons is about the speed of light $c$ [3,4].
The surface layer with a very strong 
electric field at the quark surface is similar,
in some respects, to a parallel plate
capacitor {\it which is connected to
a voltage source}. The electrons of quark matter may be considered as
such a source which gives a potential drop of $\varepsilon_{\rm F}/e
\simeq 2\times 10^7$ V across the Coulomb barrier at the quark-matter
surface [3,4], where $e$ is the electron charge. 

The energy of created electrons, $\varepsilon\leq\varepsilon_{_
{\rm F}} -2mc^2$, is smaller than the mean energy 
of electrons in the surface layer with a very strong electric 
field, $E\gtrsim E_{\rm cr}$, where the pair creation process occurs [1].
Therefore, the pair creation leads to cooling of electrons  
in the surface layer. The thermal energy which 
is stored in electrons in this layer and may be converted into
the energy of created pairs is 
$W_e\simeq 4\pi R^2\Delta R n_e(kT_{\rm S})^2/\varepsilon_{_{\rm F}}$,
where $R\simeq 10^6$ cm is the radius of the star, $n_e\simeq 3\times 
10^{34}$ cm$^{-3}$ is the density of electrons in quark matter and
$\varepsilon_{_{\rm F}}\simeq 20$ MeV [3,4]. Even if the surface 
temperature $T_{\rm S}$ is as high as $10^{11}$ K, the value of $W_e$ 
is small, $W_e\sim 10^{32}$ ergs,
i.e. the thermal energy of electrons in the surface
layer cannot support a high $e^+e^-$ emission from a strange star 
for a long time. To find the time evolution of both the 
surface temperature of a hot strange star and its luminosity in 
$e^+e^-$ pairs it is necessary to solve the
problem of heat transfer from the stellar interior to the quark-matter
surface. 

It is true that the time scale for diffusion of both electrons and photons 
from the interior of a strange star to its surface is very large,
$t_e \sim 10^{12}-10^{13}$ s.
For a very hot strange star, the energy transport can take place
on a time scale of $\sim 10$ s not only via neutrinos or other 
particles having a weak interaction cross-section, and also via 
convection. Indeed, when a hot strange star is cooling from its
surface because of the pair production, a cold dense layer forms
in the surface vicinity. This layer
is unstable with respect to convective disturbances 
[5,6] if the Rayleigh number $\tilde {\rm R} =
\alpha g \Delta Th^3/k\nu$ is 
larger than $\sim 10^3$, where $g\simeq 10^{14}$ cm s$^{-2}$ is the
acceleration of gravity at the strange-star surface, 
$k$ is the thermometric conductivity,
$\nu$ is the kinematic viscosity,
$\alpha$ is the coefficient of thermal expansion,
$\Delta T$ is the temperature change in the layer and $h$ is the
depth of the layer. At $\Delta T\simeq T\simeq 10^{11}$ K, we have [7]
$k\simeq 0.1$ cm$^2$ s$^{-1}$, $\nu \simeq 1$ cm$^2$ s$^{-1}$,
$\alpha \simeq 10^{-14}$ K$^{-1}$ and $\tilde {\rm R}\simeq 10^{12}(h/1 \,
{\rm cm})^3$. In this case, the surface layer which is
initially isothermal begins to manifest
convective behavior, $\tilde {\rm R} \gtrsim 10^3$,
in $\sim h^2/k\simeq 10^{-5}$ s when the depth of cooled quark matter
is $h\sim 10^{-3}$ cm. Since
a buoyancy force per unit volume is $g\Delta \rho= g\alpha \rho 
\Delta T$, the time-scale for the convection is roughly
$t_c\sim (R/ \alpha g\Delta T)^{1/2}\simeq 0.003$ s, here $\rho$ 
is the quark-matter density.  
This is a very rapid rate of the thermal energy transport which is able
to support a very high temperature, $T_{_{\rm S}}\gtrsim$
a few $\times 10^{10}$ K, at the surface of a young strange star
at least for a few seconds after the strange star formation. 
A detailed consideration
of both the thermal energy transport and the luminosity 
of a very hot strange star in $e^+e^-$ pairs is under way 
and will be addressed elsewhere.

\bigskip

\noindent
V.V. Usov

Department of Condensed Matter Physics

Weizmann Institute of Science

Rehovot 76100, Israel

\medskip

\noindent
Received

\noindent 
PACS numbers: 97.60.-s, 12.38.Mh

\end{multicols}

\end{document}